\documentclass[english,11pt]{article}
\pdfoutput=1
\usepackage{epsf,amsmath,amssymb,graphicx,dcolumn}
\usepackage{scalefnt,ulem}
\usepackage{subcaption}
\usepackage{cite,hyperref}
\usepackage{cleveref}
\usepackage{todonotes}
\usepackage{color}
\usepackage{rotating}
\usepackage{microtype}
\usepackage{listings}
\definecolor{lightblue}{rgb}{.7,.8,1}
\Crefname{figure}{Fig.}{Figs.}
\title{\vspace*{-6em}
  \begin{flushright}
    {\sf\small
      CERN-PH-TH-2015-080 --- ZU-TH 09/15 --- MCnet-15-08\\
    }
  \end{flushright}
\vspace*{2em}
Hadronic Higgs production through \nlo{}\plus{}\ps{}\\ in the \sm{}, the \thdm{} and the \mssm{}}
\author{Hendrik Mantler$^{a}$, Marius Wiesemann$^{b}$\\[2em]
 {\it $^a$TH Division, Physics Department, CERN,}\\[0em]
{\it CH-1211 Geneva 23, Switzerland}\\[1em]
  {\it $^b$Physik-Institut, Universit\"at Z\"urich,}\\[0em]
 {\it 8057 Z\"urich, Switzerland}\\[1em]
{\small\tt hmantler@cern.ch}\\[-.3em]
{\small\tt mariusw@physik.uzh.ch}
}
\date{}

\def\bal#1\eal{\begin{align}#1\end{align}}

\newcommand{\abbrev}{\scalefont{.9}}

\newcommand{\mhiggs}{m_{\phi}}
\newcommand{\mh}{m_{h}}

\newcommand{\mtop}{m_t}
\newcommand{\mbottom}{m_b}

\newcommand{\sushimc}{{\tt aMCSusHi}}
\newcommand{\lhc}{{\abbrev LHC}}
\newcommand{\slha}{{\abbrev SLHA}}
\newcommand{\susy}{{\abbrev SUSY}}
\newcommand{\sushi}{{\tt SusHi}}
\newcommand{\bsm}{{\abbrev BSM}}

\newcommand{\powheg}{{\abbrev POWHEG}}

\newcommand{\pysix}{{\tt Pythia6}}
\newcommand{\hwsix}{{\tt Herwig6}}
\newcommand{\pythia}{{\tt Pythia8}}
\newcommand{\hwpp}{{\tt Herwig++}}

\newcommand{\block}[1]{{\tt Block #1}}
\newcommand{\blockentry}[2]{{\tt Block #1[#2]}}

\newcommand{\pt}{\ensuremath{p_T}}

\newcommand{\nll}{{\abbrev NLL}}
\newcommand{\nnll}{{\abbrev NNLL}}
\newcommand{\lo}{{\abbrev LO}}
\newcommand{\nlo}{{\abbrev NLO}}
\newcommand{\nnlo}{{\abbrev NNLO}}
\newcommand{\sm}{{\abbrev SM}}
\newcommand{\qcd}{{\abbrev QCD}}
\newcommand{\thdm}{{\abbrev 2HDM}}
\newcommand{\mssm}{{\abbrev MSSM}}
\newcommand{\ps}{{\abbrev PS}}

\newcommand{\plus}{{\abbrev +}}
\newcommand{\citere}[1]{Ref.\cite{#1}}
\newcommand{\citeres}[1]{Refs.\cite{#1}}
\newcommand{\eqn}[1]{Eq.\,(\ref{#1})}
\newcommand{\eqns}[1]{Eqs.\,(\ref{#1})}

\newcommand{\fig}[1]{Fig.\,\ref{#1}}

\newcommand{\sct}[1]{Section~\ref{#1}}

\newcommand{\muF}{\mu_{\rm F}}
\newcommand{\muR}{\mu_{\rm R}}
\newcommand{\Qres}{\ensuremath{Q_{\text{res}}}}
\newcommand{\Qsh}{\ensuremath{Q_{\text{sh}}}}

\newcommand{\msbar}{\overline{\text{MS}}}

\newcommand{\pdf}{{\abbrev PDF}}
\newcommand{\mstw}{{\abbrev MSTW2008} $68${\abbrev \%CL}}

\newcommand{\madmc}{{\tt MadGraph5\_aMC@NLO}}
\newcommand{\mcatnlo}{{\abbrev MC@NLO}}
\newcommand{\fh}{{\tt FeynHiggs}}

\newcommand{\mhmodp}{$m_h^{\text{mod}+}$}

\newcommand{\MC}{{\tiny\mbox{MC}}}
\newcommand{\mc}{\mathcal}
\newcommand{\npo}{{n+1}}
\newcommand{\pnpo}{\Phi_\npo}
\newcommand{\pn}{\Phi_n}
\newcommand{\ppon}{\Phi_{1}}

\newcounter{notecount}
\tikzstyle{boxblue} = [draw=blue, fill=blue!10, very thick,
    rectangle, rounded corners, inner sep=5pt, inner ysep=5pt]

\begin{document}
\maketitle

\vspace*{1cm}
\begin{abstract}
\noindent
The \nlo{} cross section of the gluon fusion process is matched to parton showers in the \mcatnlo{} approach.
We work in the framework of \madmc{} and document the inclusion of the full quark-mass dependence in the \sm{} as well as the state-of-the-art squark and gluino effects within the \mssm{} embodied in the program \sushi{}. The combination of the two programs is realized by a script which is publicly available and whose usage is detailed. We discuss the input cards and the relevant parameter switches. One of our focuses is on the shower scale which is specifically important for gluon-induced Higgs production, particularly in models with enhanced Higgs-bottom Yukawa coupling.
\end{abstract}
\vfill

\thispagestyle{empty}

\clearpage
\section{Introduction}
Higgs production proceeds predominantly through gluon fusion in a large number of theories, including the Standard Model (\sm{}). The recently discovered resonance \cite{Aad:2012tfa,Chatrchyan:2012ufa} in searches for a Higgs boson is fully consistent with the \sm{} picture\footnote{See \citeres{Dittmaier:2011ti,Dittmaier:2012vm,Heinemeyer:2013tqa}
 for a theoretical overview.}, so far. Still, the measured Higgs boson may be embedded in an enlarged Higgs sector with respect to the one of the \sm{} which predicts only a single physical particle breaking the electro-weak symmetry. Two-Higgs Doublet Models (\thdm{}'s) such as the Minimal Supersymmetric \sm{} (\mssm{}) are among the most popular theories with enlarged Higgs sectors. Such theories inevitably require the existence of further physical Higgs particles. A \thdm{} predicts three neutral Higgs bosons: a light ($h$) and a heavy ($H$) scalar, and a pseudo-scalar ($A$); as well as two charged Higgs particles ($H^\pm$). Almost all \thdm{} and \mssm{} scenarios that are in agreement with the experimental bounds feature a light scalar which is \sm{}-like in its couplings to vector bosons and fermions, while the other Higgs bosons are heavier and, therefore, escaped detection up to now. Indeed, the experimental search for other Higgs resonances is one of the major focuses regarding the discovery of physics beyond the \sm{} (\bsm{}) in the second run of the Large Hadron Collider (\lhc{}).

Higgs production through gluon fusion is mediated by a colored particle. In the \sm{}, the top quark gives the dominant contribution to the cross section \cite{Georgi:1977gs,Dawson:1990zj,Djouadi:1991tka}. While also the bottom quark gives a sizable contribution, the effects due to other quarks are small and therefore usually neglected. In the \thdm{} and the \mssm{} the Higgs-bottom Yukawa coupling can be enhanced with respect to the one of the top quark and the bottom loop may even constitute the dominant contribution to the cross section. In those models it is stringently required to include the bottom-quark contribution.\footnote{Note that in theories with an enhanced bottom Yukawa the associated Higgs production with bottom quarks becomes relevant, see \citeres{Campbell:2002zm,Harlander:2003ai,Dittmaier:2003ej,
Dawson:2003kb,Harlander:2010cz,Harlander:2011fx,Buehler:2012cu,Harlander:2014hya,Wiesemann:2014ioa} and references therein.} The gluon fusion cross section is known at the next-to-leading order (\nlo{}) in the \sm{} including top- and bottom-mass effects \cite{Spira:1995rr,Harlander:2005rq}, in the \thdm{} and in the \mssm{} including contributions from squarks and gluinos \cite{Harlander:2003bb,Harlander:2004tp,Harlander:2003kf,Harlander:2005if,Muhlleitner:2006wx,Aglietti:2006tp,Bonciani:2007ex,Degrassi:2008zj,Anastasiou:2008rm,Harlander:2010wr,Degrassi:2010eu,Muhlleitner:2010nm,Degrassi:2011vq,Bagnaschi:2011tu,Degrassi:2012vt}. For the top quark, an effective field theory approach can be applied in which the top quark is considered to be infinitely heavy and can be integrated out from the full theory. In this approximation, Higgs production has been calculated up to the next-to-\nlo{} (\nnlo{}) inclusively \cite{Harlander:2002wh,Anastasiou:2002yz,Ravindran:2003um} as well as fully-differential \cite{Anastasiou:2004xq,Catani:2007vq,Grazzini:2008tf}. Electro-weak contributions and effects beyond \nnlo{} in the heavy-top approximation have been studied in \citeres{Catani:2003zt,Idilbi:2005ni,Ravindran:2006cg,Ahrens:2008nc,Djouadi:1994ge,Degrassi:2004mx,Anastasiou:2008tj,Aglietti:2004nj,Actis:2008ug,Bonciani:2010ms} for example, while there was a large effort \cite{Anastasiou:2014vaa,Anastasiou:2014lda} to push the accuracy to next-to-\nnlo{} ({\abbrev N$^3$LO}) which has been succeeded very recently \cite{Anastasiou:2015ema}. Finite top-mass effects have been shown to be small for both the inclusive cross section at the \nnlo{} \cite{Marzani:2008az,Harlander:2009bw,Harlander:2009mq,Harlander:2009my,
Pak:2009dg,Pak:2011hs} and differential quantities \cite{Harlander:2012hf,Neumann:2014nha} as long as no kinematical scale (such as the transverse momentum of a particle) that is not integrated out exceeds the top-mass threshold.

The full dependence of the top- and the bottom-mass at the \nlo{} has been included so far in a \powheg{}-type \cite{Nason:2004rx} matching to parton showers (\ps{}s) \cite{Bagnaschi:2011tu}, the analytically resummed transverse momentum spectrum of the Higgs boson at \nlo{}\plus{}\nll{} \cite{Mantler:2012bj}, a \mcatnlo{}-type \cite{Frixione:2002ik} matching to the {\tt Herwig} Monte Carlos \cite{MCNLO410,Frixione:2010wd,Frixione:2010ra}, the \nnlo{}\plus{}\nnll{} jet-vetoed \cite{Banfi:2013eda} and the fully-differential \nnlo{} \cite{Grazzini:2013mca} cross section; and in some approximated form recently also  in the \nnlo{}\ps{} approach \cite{Hamilton:2013fea,Hamilton:2015nsa}. Furthermore, the \thdm{} as well as supersymmetric effects from squarks and gluinos within the \mssm{} \cite{Harlander:2003bb,Harlander:2004tp,Harlander:2003kf,Harlander:2005if,Muhlleitner:2006wx,Aglietti:2006tp,Bonciani:2007ex,Degrassi:2008zj,Anastasiou:2008rm,Harlander:2010wr,Degrassi:2010eu,Muhlleitner:2010nm,Degrassi:2011vq,Bagnaschi:2011tu,Degrassi:2012vt} have been implemented in the first two approaches from that list \cite{Bagnaschi:2011tu,Harlander:2014uea}. In this manuscript, we report on a new implementation of \nlo{} \qcd{} corrections in the \sm{}, \thdm{} and \mssm{} applying the \mcatnlo{}-type matching to both {\tt Herwig} and {\tt Pythia} showers. We work in the framework 
of \madmc{} \cite{Alwall:2014hca} and combine its capabilities with the corresponding amplitudes provided by \sushi{} \cite{Harlander:2012pb}. The linking of \sushi{} to \madmc{} is realized by a script\footnote{\sushimc{} can be downloaded under  \url{https://cp3.irmp.ucl.ac.be/projects/madgraph/wiki/HSushi\#no1}.} \sushimc{}. Its usage as well as the application of the combined code to obtain cross section predictions in the \sm{}, the \thdm{} and the \mssm{} is detailed in this paper.\footnote{\sushi{} has recently been extended to include {\abbrev NMSSM} Higgs production \cite{Liebler:2015bka}, which may be made available in \sushimc{} in the future.}

The manuscript is organized as follows: In \sct{sec:outline} we present a brief overview of the elements of the computation at hand. \sct{sec:script} is dedicated to introduce \sushimc{} and is separated in three parts which cover: how to use the script (\sct{sec:use}), how to run the resulting code (\sct{sec:run}) and how to treat the shower scale (\sct{sec:scales}). We will show a brief application 
of the code to phenomenological results in \sct{sec:results} and conclude in \sct{sec:conclusions}.

\section{Outline of the calculation}
\label{sec:outline}
The goal of this paper is to present a tool which allows for the computation of arbitrary infra-red safe 
differential observables at both the parton- and the hadron-level for the production of neutral 
Higgs bosons via gluon fusion in the \sm{}, the \thdm{} and the \mssm{} by matching the
\nlo{} cross section to a shower. 

\begin{figure}[t]
  \begin{center}
    \begin{tabular}{ccccc}
      \mbox{\includegraphics[height=.12\textheight]{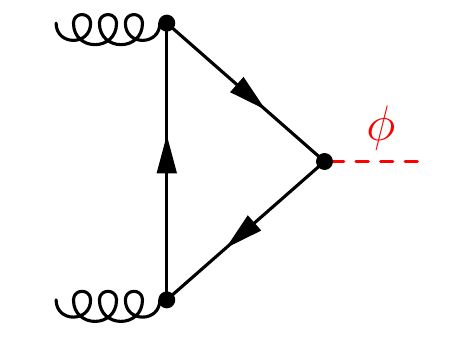}} & & \mbox{\includegraphics[height=.12\textheight]{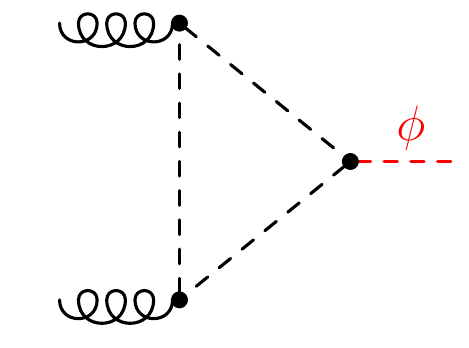}} & &
      \mbox{\includegraphics[height=.12\textheight]{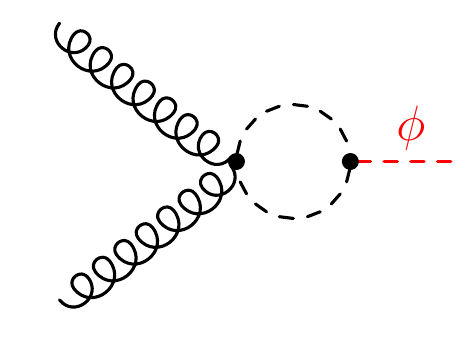}}
            \\
      (a) & & (b) & & (c)\\
      \mbox{\includegraphics[height=.12\textheight]{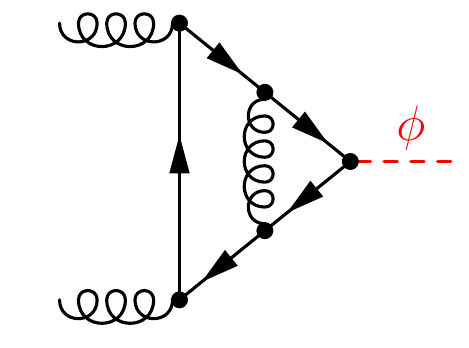}} & & \mbox{\includegraphics[height=.12\textheight]{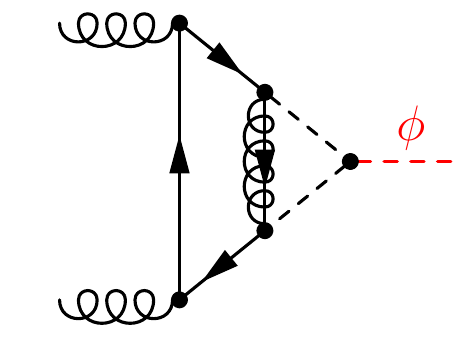}} & &
      \mbox{\includegraphics[height=.12\textheight]{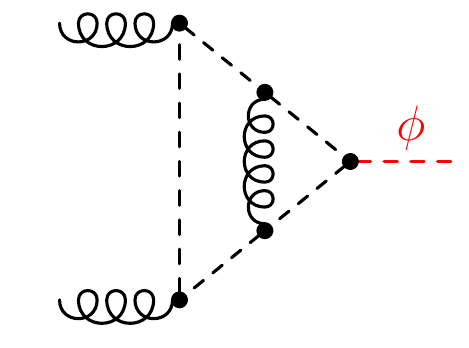}}
            \\
      (d) & & (e) & & (f)\\
      \mbox{\includegraphics[height=.12\textheight]{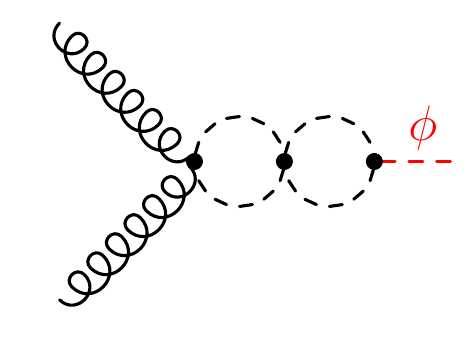}} & & \mbox{\includegraphics[height=.085\textheight]{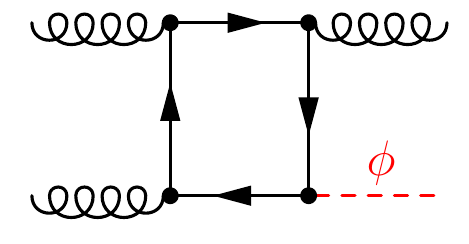}} & &
      \mbox{\includegraphics[height=.11\textheight]{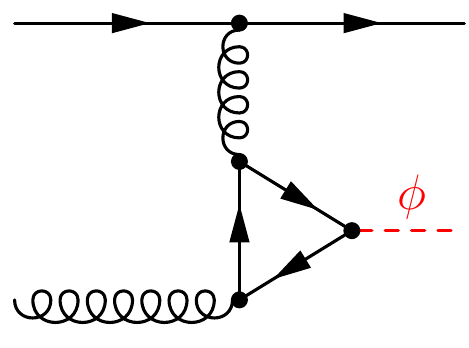}}
            \\
      (g) & & (h) & & (i)
    \end{tabular}
    \parbox{.9\textwidth}{%
      \caption[]{\label{fig:diag} A sample of Feynman diagrams for
        $gg\rightarrow \phi$ contributing to the \nlo{} cross section;
        (a-c) \lo{}, (d-g) virtual and (h-i) real corrections. The
        graphical notation for the lines is: solid straight
        $\widehat{=}$ quark; spiraled $\widehat{=}$ gluon; dashed
        $\widehat{=}$ scalar (squark or Higgs); spiraled with line
        $\widehat{=}$ gluino.  } }
  \end{center}
\end{figure}

The relevant \nlo{} matrix elements are taken from \citere{Harlander:2012pb}, which include both
\sm{}-like contributions and sbottom, stop and gluino effects. Examples of corresponding Feynman 
diagrams are illustrated in \fig{fig:diag}. They are combined and matched to a parton shower by the 
well-known \mcatnlo{}-method. The matched cross section 
in the \mcatnlo{} framework can be written symbolically as:
\begin{align}
\begin{split}\label{eq:mcatnlo}
\left(\frac{d\sigma}{dO}\right)_{\text{MC@NLO}}=&\int d\pn \left[B_n+V_n+\int d\ppon^\MC\,K^\MC_\npo\right]\mc I^\MC_n(O)\\
&+\int \left[d\pnpo R_\npo -d\pnpo^{\MC} K^\MC_\npo\right]\mc I^\MC_\npo(O)\,,
\end{split}
\end{align}
where $B_n$ determines the Born-level cross section, $V_n$ the virtual (including mass factorization) and $R_\npo$ the real corrections; $K^\MC_\npo$ is the Monte Carlo subtraction term, with the same {\abbrev IR} poles as $R_\npo$, the Monte Carlo phase space $d\pnpo^\MC$ tends to $d\pnpo$ in the {\abbrev IR} limits, and $d\ppon^\MC=d\pnpo^\MC/d\pn$. The quantity $\mc I_n^\MC(O)$ is the shower spectrum for observable $O$, as obtained by running the shower starting from an $n$-body configuration.

The cross section at Born-level is derived from the \lo{} diagrams for $gg\rightarrow \phi$ where $\phi\in\{h,H,A\}$, see e.g. \fig{fig:diag}\,(a)-(c). 
The \nlo{} virtual and real corrections are governed by diagrams like the ones shown in \fig{fig:diag}\,(d)-(g) and \fig{fig:diag}\,(h)-(i), respectively, and similar ones with
quark loops replaced by squark loops.

\eqn{eq:mcatnlo} is implemented for all standard parton showers \cite{Corcella:2000bw,Corcella:2002jc,Sjostrand:2006za,Sjostrand:2007gs,Bahr:2008pv,Bellm:2013lba} in the fully-automated framework \madmc{}. This code determines
 \nlo{} \qcd{} corrections to arbitrary scattering processes at the \lhc{}. On the basis of {\abbrev UFO} models \cite{Degrande:2011ua}, 
the code even allows to carry out computations in any theory beyond the \sm{} in a general manner as soon as the renormalization 
is known and implemented in a {\abbrev UFO} model, see \citeres{Christensen:2008py,Christensen:2009jx,Alloul:2013bka,Alwall:2014hca} for further information. However, the Higgs production mode 
 through gluon fusion is a special one being loop-induced already at the \lo{}. Such processes cannot be handled 
in a fully-automated manner by any code to date, since it requires the automation of two-loop amplitudes which is beyond current 
technology. Therefore, we have treated Higgs production through gluon fusion in the \sm{}, \thdm{} and \mssm{} as a special 
case, by linking the relevant amplitudes from \sushi{}. Furthermore, as far as the \mssm{} is concerned \sushi{} requires a link to {\tt FeynHiggs} \cite{Heinemeyer:1998yj,Heinemeyer:1998np,Degrassi:2002fi,Frank:2006yh,Hahn:2013ria,Degrassi:2001yf,Brignole:2001jy,Brignole:2002bz,Dedes:2003km,Heinemeyer:2004xw,Heinemeyer:2007aq} 
 which evaluates the corresponding Higgs masses and couplings in user defined scenarios. 
Setting up the \sushi{} amplitudes in \madmc{} is handled by a publicly available script called \sushimc{}, which is automated to create the $gg\rightarrow\phi$ process folder; download \sushi{} and  {\tt FeynHiggs};
compile, install and link them; and replace the relevant amplitudes in the process folder. In the upcoming section, we describe the application 
of the script and explain the necessary steps to obtain phenomenological results.

\section{\sushimc{} script}
\label{sec:script}
This code is based on \madmc{} and \sushi{}. For further information on theses codes we refer the reader to corresponding publications \cite{Alwall:2014hca,Harlander:2012pb}. After using the script to set up the code, we will focus on the relevant user inputs to obtain phenomenological predictions for Higgs cross sections in the \sm{}, the \thdm{} and the \mssm{}.

\subsection{Usage of the script}
\label{sec:use}

The \sushimc{} script is available for download from the website \url{https://cp3.irmp.ucl.ac.be/projects/madgraph/wiki/HSushi\#no1}. It is fully automatic in setting up the $gg\rightarrow\phi$ process folder which includes downloading, installing and linking \fh{} and \sushi{}. At first, a dummy {\abbrev HEFT} process folder for $gg\rightarrow h$ in the five-flavor scheme at \nlo{} (without the virtuals) is created;\footnote{See the {\abbrev README} file how this can be easily obtained with the \madmc{} script.} then, the {\abbrev HEFT} amplitudes in \madmc{} are consistently replaced by the ones from \sushi{} (including the virtuals). The set-up requires only a single call of the \sushimc{} script:

\begin{tikzpicture}
\node[text width=0.98\textwidth,boxblue] {
 $>$ ./set\_up\_ggH\_MSSM\_script.pl $<$ggH-folder$>$ [$<$\fh{}-folder$>$ [$<$\sushi{}-folder$>$]]
};
\end{tikzpicture}

The first argument is mandatory and determines the path to the process folder for $gg\rightarrow\phi$ that is generated by the script. The only requirement is that this folder has to be defined as a sub-folder of the \madmc{} directory. The second and third arguments are optional. If there are compiled versions of \fh{} and \sushi{} available on your computer you can give the names of the folders that contain the files libFH.a and libsushi.a, respectively. When executed with two (one) arguments the script will ask whether it should automatically download and install \sushi{} (and \fh{}). The script will always download the latest versions of these codes. While running, the script requires some user inputs: It asks whether or not \sushi{} (and \fh{}) should be downloaded, in which folder they should be installed (default is inside the $<$ggH-folder$>$) or where to find \sushi{} (and \fh{}) if already installed. The user is simply required to follow these on-screen instructions. Furthermore, the script creates log-files in the working directory for the download ("XX\_curl.log``), the configure command ("XX\_conf.log``) and the compilation ("XX\_make.log``), where XX$=$FH for \fh{} and XX$=$SusHi for \sushi{}. These files are supposed to give complementary information for any kind of troubleshooting.\footnote{Please note that even in cases where the compilation of \fh{} or \sushi{} fails, the linking may still work fine as long as the files libFH.a and libsushi.a have been created in the corresponding library folders.} Further information about the \sushimc{} script provides the {\abbrev README} file.

\subsection{Running the code}
\label{sec:run}
Once the $gg\rightarrow\phi$ process folder has been set up by the script, the run can be started directly from the $<$ggH-folder$>$ by typing

\begin{tikzpicture}
\node[text width=0.98\textwidth,boxblue] {
$>$ ./bin/generate\_events
};
\end{tikzpicture}
and following the usual steps in \madmc{} to choose the run-modes (order, shower or fixed-order, madspin). Before running the code 
one may want to modify the input settings. In the following we will discuss the differences between \sushimc{} and the ordinary \madmc{} code regarding the input files.

The input cards can be found under $<$ggH-folder$>$/Cards/, where the param\_card.dat, run\_card.dat and shower\_card.dat contain 
all the essential information. The run card controls the usual parameters, e.g. the renormalization ($\muR$) and factorization scale ($\muF$). Note, that by default the flags {\tt fixed\_ren\_scale} and {\tt fixed\_fac\_scale} are set to false, so that these two scales are chosen on an event-wise basis. Their values are specified in the last routine of $<$ggH-folder$>$/SubProcesses/setscales.f which, due to the default option ${\tt dynamical\_scale\_choice}=-1$ in the run card, sets $\muR=\muF=H_T/2\equiv1/2\,\sum_{i}(m_i^2+p_T^2(i))^{1/2}$, where $i$ runs over all final state particles and $m_i$ and $p_T(i)$ are their mass and transverse momentum, respectively. This choice is reasonable, since it respects effects from hard radiation and corresponds to a value of $\mhiggs{}/2$ in the soft/collinear limit which is the current recommendation for the total inclusive $gg\rightarrow\phi$ cross section \cite{Dittmaier:2011ti,Heinemeyer:2013tqa}.

Also the shower card in \sushimc{} contains no new information and has the usual functionality. Since \madmc{} supports all standard parton showers, for the first time \pysix{} and \pythia{} can be applied at \nlo{}\plus{}\ps{} to \sm{} Higgs  production in the full theory in the \mcatnlo{} framework. In general, it is advisable to apply the most recent versions of the showers for meaningful physics runs.

The param\_card.dat, on the other hand, receives some significant changes by the \sushimc{} script. The new version basically combines the orignial parameter card from \madmc{} with the input file from \sushi{} which are both written in the \susy{} Les Houches accord (\slha{}) format \cite{Skands:2003cj} and, therefore, easily connectable. We will address the different options in the param\_card.dat in more detail, since there are a number of changes and some of the original parameters loose their functionality. In the \slha{} format inputs are organized in blocks which have different entries that are characterized by a number. For simplicity, we introduce the following short-hand notation: \blockentry{example}{i} corresponds to entry $i$ in \block{example}. E.\,g., entry 25 of \block{mass} (\blockentry{mass}{25}) in the \slha{} format is devoted to the Higgs mass in the \sm{}, which is required as an input in the param\_card.dat. A typical parameter card of \sushimc{} in the \sm{} is shown below:

\lstset{basicstyle=\scriptsize, frame=shadowbox}
{\tt
\begin{lstlisting}
###################################
## INFORMATION FOR MASS
###################################
Block mass 
   15 1.777000e+00 # MTA
   23 9.118800e+01 # MZ
   25 1.250000e+02 # MH -- only effective if FEYNHIGGS Block is absent
[...]

###################################
## INFORMATION FOR SMINPUTS
###################################
Block sminputs 
    1 1.325070e+02 # aEWM1 
    2 1.166390e-05 # Gf 
    3 1.180000e-01 # aS 
# additional information needed for SusHi
    4 9.118760e+01 # m_Z(pole)
    5 0.416000e+01 # m_b(m_b)  -- only used if m_b is not on-shell
    6 1.730000e+02 # m_t(pole) -- top mass is set here

###################################
## INFORMATION FOR YUKAWA
###################################
Block yukawa 
   15 1.777000e+00 # ymtau 

###################################
## INFORMATION FOR DECAY
###################################
DECAY   6 1.491500e+00 # WT 
DECAY  23 2.441404e+00 # WZ 
DECAY  24 2.047600e+00 # WW 
DECAY  25 6.382339e-03 # WH 
DECAY 9000006 6.382339e-03 # WH1 
[...]

###################################
## INFORMATION FOR SUSHI
###################################
Block sushi
  1   0		# model: 0 = SM, 1 = MSSM, 2 = 2HDM
  2   0         # 0 = light Higgs (h), 1 = pseudoscalar (A), 2 = heavy Higgs (H)
Block renormbot # Renormalization of the bottom sector
  1   0 	# m_b used for bottom Yukawa:  0 = OS, 1 = MSbar(m_b), 2 = MSbar(muR)
  4   4.75d0	# mbOS fixed -- used if m_b is on-shell (default)
Block factors
  1   0.d0	# factor for yukawa-couplings: c
  2   1.d0	# t
  3   1.d0	# b
\end{lstlisting} }

Most of the inputs are self-explanatory due to the comments initiated by the hash symbol {\tt \#} after the entries. Furthermore, the standard \slha{} blocks match the universal convention of \citere{Skands:2003cj}. Some of the inputs, though, require further comments. In the \block{mass} all parameters have the expected function, except for the top and the bottom mass, \blockentry{mass}{6} and \blockentry{mass}{5}, respectively. While the former only affects and is required for the shower, the latter can be omitted completely. Instead, due to the link to \sushi{} the top mass that is used for the top loop and the top Yukawa is set in \blockentry{sminputs}{6} and is expected to be on-shell. For the bottom mass, on the other hand, \sushi{} allows for three different choices: on-shell scheme or $\msbar{}$ scheme with $m_b(m_b)$ or $m_b(\muR)$, which can be switched in \blockentry{renormbot}{1} by a value between $1$ and $3$.\footnote{Note that only in the on-shell scheme the automatic \madmc{} reweighting for $\muR{}$ is functionable.} Also here the recommendation is to use the on-shell scheme which according to \citeres{Spira:1995rr,Bagnaschi:2014zla} assures the cancellation of large logarithms $\ln(\mh/\mbottom)$ at \nlo{} \qcd{}, while the $\msbar{}$ scheme does not, due to an incomplete resummation of these terms. The on-shell $\mbottom$ value is determined by \blockentry{renormbot}{4}, while when the $\msbar{}$ scheme is chosen the corresponding input of $\mbottom(\mbottom)$ is set in \blockentry{sminputs}{5}. The other entries of \block{sminputs} again have the same impact as in the usual \madmc{} code. The same is true for \block{yukawa}. For the decay of light and heavy Higgs bosons one may specify a finite width of the Higgs boson in the respective \bsm{} scenario by using {\tt Decay 25} irrespective of wether the light or the heavy Higgs boson is considered. At this point we shall remark that the particle identification number ({\abbrev PID}) in the generated event files is always {\tt 25} regardless of the Higgs boson under consideration.\footnote{Bear in mind that this has to taken into account to identify the Higgs in the analysis of the showered events.} This is irrelevant for the production (which is correctly computed through the \sushi{} amplitudes), but plays a role for the decay where the shower will consider particle {\tt 25} to be the light Higgs, which is indeed fine for any scalar Higgs, but a problem for pseudo-scalar ones. Therefore, decays of a pseudo-scalar Higgs are currently not supported in the official version of \sushimc{}. A user interested in decaying the pseudo-scalar Higgs is strongly encouraged to contact us.

The other parameters are relevant to \sushi{}. \blockentry{sushi}{1} chooses the model with the three options \sm{} (\blockentry{sushi}{1}$=0$), \mssm{} (\blockentry{sushi}{1}$=1$) and \thdm{} (\blockentry{sushi}{1}$=2$). The second entry of \block{sushi} determines the Higgs boson: light Higgs (\blockentry{sushi}{2}$=11$ or $0$), pseudo-scalar Higgs (\blockentry{sushi}{2}$=21$ or $1$) and heavy Higgs (\blockentry{sushi}{2}$=12$ or $2$). The choice of the masses of the relevant Higgs bosons depends on the model. As stated before, \blockentry{mass}{25} sets the Higgs mass in the \sm{}. In the \thdm{}, this entry corresponds to the mass of the light Higgs boson, while \blockentry{mass}{35} and \blockentry{mass}{36} specify the input for the heavy and the pseudo-scalar Higgs, respectively. All other \thdm{} inputs are set in the information for \sushi{}. For reference, we give an example of the corresponding inputs for a heavy Higgs in the \thdm{} below:\footnote{A link to {\tt 2HDMC} \cite{Eriksson:2009ws} with the corresponding input convention is currently not supported.}

\lstset{basicstyle=\scriptsize, frame=shadowbox}
{\tt
\begin{lstlisting}
[...]

###################################
## INFORMATION FOR SUSHI
###################################
Block sushi
  1   2		# model: 0 = SM, 1 = MSSM, 2 = 2HDM
  2   2         # 0 = light Higgs (h), 1 = pseudoscalar (A), 2 = heavy Higgs (H)
Block renormbot # Renormalization of the bottom sector
  1   0 	# m_b used for bottom Yukawa:  0 = OS, 1 = MSbar(m_b), 2 = MSbar(muR)
  2   2 	# tan(beta)-res. of Y_b: 0 = no, 1 = naive, 2 = full (for OS only)
  4   4.75d0	# mbOS fixed
Block 2hdm # 2HDM version according to arXiv:1106.0034
  2             # (1=I,2=II ,3=III ,4=IV)
Block minpar
  3   50d0      # tanb
Block alpha
  0.0247d0      # mixing in Higgs sector
Block factors
  1   0d0       # factor for yukawa-couplings: c
  2   1d0       # t
  3   1d0       # b
\end{lstlisting} }

Additionally to the inputs which we defined already for the \sm{} the following parameters have to be set in the \thdm{}: \blockentry{renormbot}{2} specifies whether or not a resummation\footnote{See \citeres{Carena:1999py,Guasch:2003cv,Noth:2010jy,Noth:2008tw,Hofer:2009xb,Mihaila:2010mp} for further information.} of terms enhanced by $\tan(\beta)$ is applied through reweighting of the bottom Yukawa coupling as described in \citere{Harlander:2012pb}; \block{2hdm} determines which type of the \thdm{} is used; the value of $\tan(\beta)$ is set through \blockentry{minpar}{3}; and the mixing angle $\alpha$ corresponds to the entry in \block{alpha}.

The computation of \mssm{} Higgs cross sections requires \block{extpar}, \block{feynhiggs} and \block{renormsbot} in addition, which fix the parameters of the third family of quarks and squarks, determine the \fh{} inputs and yield information on the renormalization of the sbottom section, respectively. We will not provide any further information on these blocks, instead, we refer to the \sushi{} manual \cite{sushimanual} and the \fh{} man pages \cite{fhmanual}. Moreover, \block{alpha} can be omitted in the \mssm{} and the Higgs masses in \block{mass} have no effect, since they are determined by \fh{}, once \block{feynhiggs} is present. The \mssm{} Higgs mass that has been computed and applied in the run is provided to the user in \blockentry{mass}{25} of the parameter card, which will be overwritten by the mass of the respective Higgs boson at the beginning of each \mssm{} run.

So far we did not comment on the \block{factors}. It allows to turn on and off individual contributions in all models. In fact, it even provides the possibility to rescale the respective Yukawa couplings by choosing values different from zero and one. With \blockentry{factors}{1} one can include the charm quark in the computation. This requires to specify its $\msbar{}$ mass $m_c(m_c)$ in \blockentry{sminputs}{8} which is then translated to its on-shell mass. Furthermore, \blockentry{factors}{2} and \blockentry{factors}{3} multiply the top and the bottom Yukawa, respectively. In the \mssm{}, the stop Yukawa is rescaled by \blockentry{factors}{4} and the sbottom one by 
\blockentry{factors}{5}.

For further information on the input cards we refer the reader to \citere{Alwall:2014hca} of \madmc{}, the manual of \sushi{} \cite{sushimanual} and the man pages of \fh{} \cite{fhmanual}. Three example parameter cards are provided in the folder $<$ggH-folder$>$/Cards; one for the \sm{} (param\_card.dat\_SM), the \thdm{} (param\_card.dat\_2HDM\_scenB) and the \mssm{} (param\_card.dat\_MSSM\_mhmodp). They match the scenarios that we study in the result section of this paper.

\subsection{Choosing different shower scales}
\label{sec:scales}

The choice of the shower scale is a very peculiar one in the gluon fusion process. In presence of the bottom-quark loop, factorization of soft and collinear radiation maybe spoiled at scales significantly smaller than the Higgs boson mass. This was pointed out by \citere{Grazzini:2013mca} in the context of analytic transverse momentum resummation. On the other hand, these terms might well be treated as a finite remainder as long as their impact remains moderate \cite{Banfi:2013eda}.

Due to their additive matching of the resummed low-\pt{} region with the fixed-order distribution valid at large transverse momenta, analytic \pt{}-resummation and the \mcatnlo{} method are quite similar. In both cases there is a scale associated with that matching, the resummation scale \Qres{} and the shower scale \Qsh{}, respectively. These scales can be interpreted as transition scales that separate the soft/collinear from the hard physics, very similar to the factorization scale of the \pdf{}s. In other words, they define the range where resummation, and therefore the shower in \mcatnlo{}, takes effect. Their value has to be chosen of the order of the typical scale of the problem.

In gluon fusion, the typical scale depends on the quark considered in the loop. Since $\mtop{}\sim\mhiggs$, there exist only two relevant scales for the top-quark loop ($\mhiggs$ and $\pt{}$) and the shower scale can be chosen of the order of the Higgs mass. When considering the bottom loop, on the other hand, we face a three-scale problem ($\mhiggs$, $\mbottom$ and $\pt{}$) which has not been solved to date. However, it has been suggested \cite{Grazzini:2013mca} to apply a lower transition scale to the bottom contribution, which, in particular, respects the fact that soft/collinear factorization is valid only up to smaller scales for the bottom loop. In \citere{Harlander:2014uea} it was further proposed to separate three contributions according to their Yukawa couplings: the square of the top and the bottom, and their interference; and choose separate shower/resummation scales for all of them. This splitting allows for a model independent treatment of the problem by a rescaling of the individual contributions with the respective top  and bottom Yukawas of a specific scenario in the \thdm{} as well as the \mssm{} when neglecting squark effects. In the literature, two pragmatic approaches with physical motivation have been presented \cite{Harlander:2014uea,BVpaper} to determine separate scales for the three contributions. Their comparison will be studied elsewhere \cite{BHMVWpaper}. When studying phenomenological results in \sct{sec:results} we will apply the
scales from \citere{Harlander:2014uea} (referred to as "{\abbrev HMW}`` in what follows).

The separation of the bottom contribution (including the interference) from the top one with different shower scales ($Q_b$ and $Q_t$, respectively) requires three runs in \madmc{}, which have to be combined as follows:
\bal
\label{eq:twoscales}
\sigma(Q_t,Q_b)=\sigma_t(Q_t)+\sigma_{t+b}(Q_b)-\sigma_t(Q_b).
\eal
To obtain all three contributions of different Yukawa origin with different scales that allows for a model independent treatment, on the other hand, \madmc{} has to be run five times:
\bal
\label{eq:threescales}
\sigma(Q_t,Q_b,Q_{tb})=\sigma_t(Q_t)+\sigma_b(Q_b)+\sigma_{t+b}(Q_{tb})-\sigma_t(Q_{tb})-\sigma_b(Q_{tb}).
\eal
The scales $Q_t$, $Q_b$ and $Q_{tb}$ determine the scale for top, the bottom and their interference, respectively. As indicated before, the individual contributions can be separated using the \block{factors} in the parameter card.

The shower scale in \madmc{} cannot simply be accessed through the input cards, since it requires an advanced user to be familiar with its specific treatment in the code. \madmc{} does not use a simple fixed scale for $\Qsh$, instead, it statistically extracts the shower scale from a distribution peaked at a specific value. The user can only change the range of the interval of the distribution which of course also affects the peak. Therefore, we identify $Q_t$, $Q_b$ and $Q_{tb}$ in \eqns{eq:twoscales} and \eqref{eq:threescales} with the peak of the respective shower scale distributions.

The so-called shape parameters define the interval of the distribution from which the shower scale is picked on a event-wise basis. They can be specified in the include file $<$ggH-folder$>$/SubProcesses/\\madfks\_mcatnlo.inc, where the relevant part is given by (\madmc{} default values):

\lstset{basicstyle=\scriptsize, frame=shadowbox}
{\tt
\begin{lstlisting}
[...]

c Define lower and upper veto range (see MC subtraction terms)
      double precision frac_low,frac_upp
      parameter (frac_low=0.1d0)
      parameter (frac_upp=1.0d0)

c Other control switches on veto (see MC subtraction terms)
      double precision scaleMClow,scaleMCdelta
      parameter (scaleMClow=10.d0)
      parameter (scaleMCdelta=20.d0)

[...]
\end{lstlisting} }

The parameters {\tt frac\_low} and {\tt frac\_upp} are used to compute the upper and lower bounds of the \Qsh{} distribution, which will be explained in more detail below, while {\tt scaleMClow} allows to set an absolute value of the lower bound on \Qsh{} and {\tt scaleMCdelta} is used to apply a minimal value to the size of the distribution interval. In formulas the interval is defined by\footnote{For further information we refer the reader to Section 2.4.4 of \citere{Alwall:2014hca}.}
\bal
\begin{split}
&Q_{\text{min}}\le \Qsh\le Q_{\text{max}}, \;\;\text{ with} \\
&Q_{\text{min}}=\max({\tt frac\_low}\cdot\sqrt{s_0},{\tt scaleMClow}) \;\;\text{ and }\\
&Q_{\text{max}}=\max({\tt frac\_upp}\cdot\sqrt{s_0},Q_{\text{min}}+{\tt scaleMCdelta}),
\end{split}
\eal
where $s_0$ is the Born-level partonic center of mass energy squared. Evidently, {\tt scaleMClow} and {\tt scaleMCdelta} only take effect if the interval obtained through {\tt frac\_low} and {\tt frac\_upp} does not meet the corresponding restrictions. The corresponding $\Qsh$ distribution is peaked around
\bal
Q_{\text{peak}}\sim\frac{({\tt frac\_low}+{\tt frac\_upp})}2\,\sqrt{\langle s_0\rangle}.
\eal
For a $2\rightarrow 1$ process like gluon fusion this relation is an identity and $\sqrt{s_0}$ equals the mass of the final state particle, i.\,e., the Higgs mass in the case of gluon fusion. To change the peak-value to its half, e.\,g., for shower scale variations, one can simply divide {\tt frac\_low} and {\tt frac\_upp} by a factor of two. In this sense, it is convenient to keep the ratio between {\tt frac\_low} and {\tt frac\_upp} a constant, which in the default setup of \madmc{} is a factor of ten. Under this prerequisite, in order to choose a specific shower scale \Qsh{} for $gg\rightarrow \phi$ we simply have to determine
\bal
{\tt frac\_upp}= \frac{2}{1.1}\cdot\frac{\Qsh}{\mhiggs} \quad\text{and}\quad {\tt frac\_low}=\frac{{\tt frac\_upp}}{10}.
\eal
Here and in what follows, we associate $Q_{\text{peak}}$ with the shower scale \Qsh{} and vice versa unless stated otherwise. After modifying the corresponding include file accordingly, \madmc{} has to be recompiled. This can be achieved by typing 

\begin{tikzpicture}
\node[text width=0.98\textwidth,boxblue] {
$>$ make clean
};
\end{tikzpicture}

inside the $<$ggH-folder$>$ which forces a recompilation during the next run of the code.

In \sct{sec:results} we show some applications of the \sushimc{} code and study the effect of different treatments of the shower scales.

\section{Results: brief application}
\label{sec:results}

The $gg\rightarrow \phi$ process folder created by the \sushimc{} script preserves all the highly convenient features that come with \madmc{}. Besides many others, this entails an interface to the most common showers, the fully-automatic determination of scale and \pdf{} variations without any extra-costs of computing time \cite{Frederix:2011ss}, the creation of any number of observables with a single run and analysis routines available for the most important processes including the gluon fusion Higgs production mode.

\begin{figure}[t]
\begin{center}
    \includegraphics[width=0.65\textwidth]{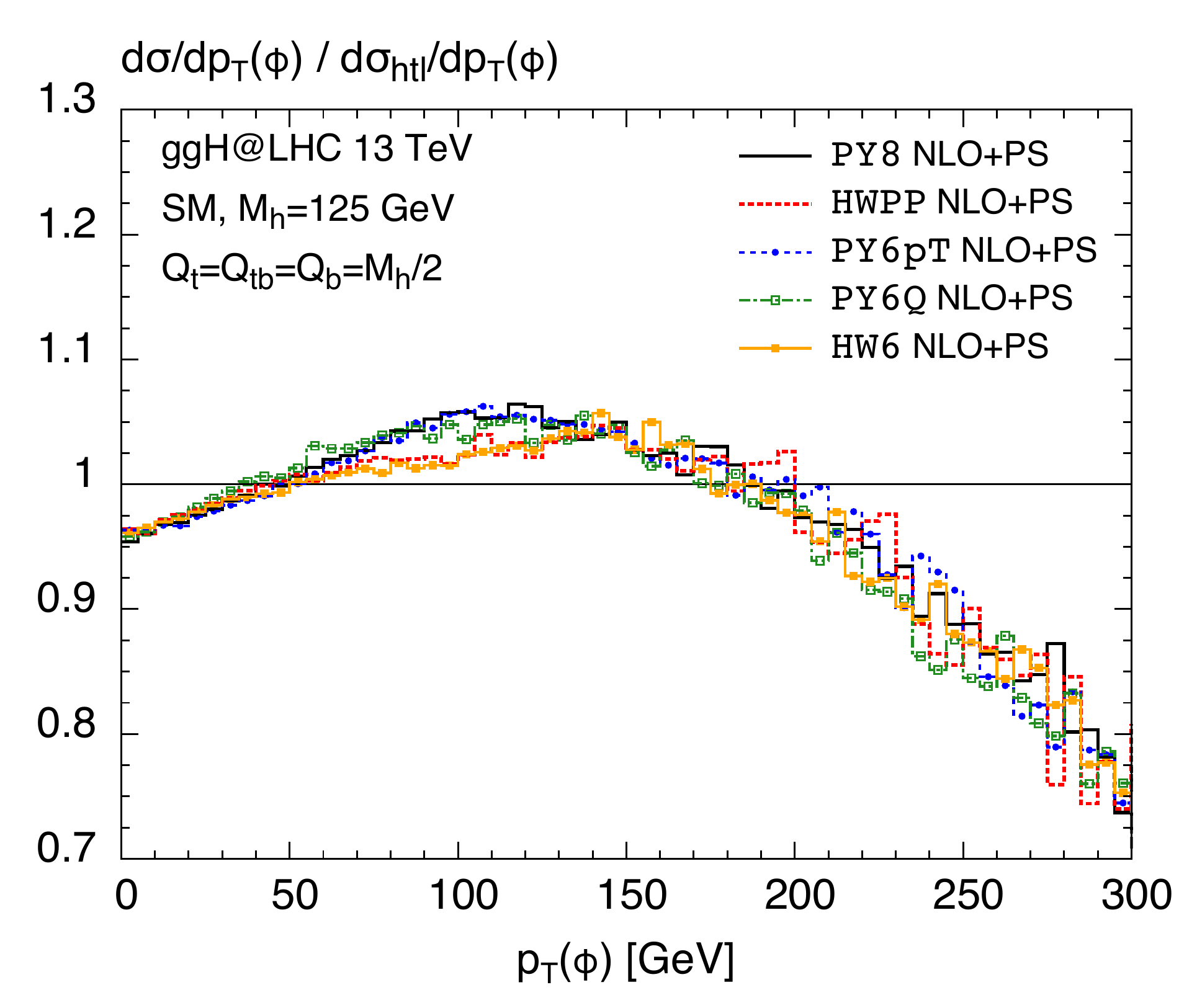}
	\parbox{.9\textwidth}{%
    \caption{\label{fig:me} Transverse momentum distribution of a \sm{} Higgs at \nlo{}\plus{}\ps{} in the full theory normalized to the one in the heavy top effective field theory for different Monte Carlos: \pythia{} (black solid), \hwpp{} (red, dotted), \pysix{} $\pt{}$-ordered (blue, dashed with points), \pysix{} $Q$-ordered (green, dash-dotted with open boxes) and \hwsix{} (yellow, solid with filled boxes).}}
\end{center}
\end{figure}

\sushimc{} allows to compute gluon-induced Higgs production including the complete dependence on the quark masses in the \sm{} for the first time in a \mcatnlo{}-type matching applying all versions of {\tt Pythia} and {\tt Herwig} Monte Carlos. While previous computations did only feature the {\tt Herwig} showers \cite{MCNLO410,Frixione:2010wd,Frixione:2010ra}, phenomenological results exist to our knowledge only for \hwsix{} \cite{Frixionetalk}. As a first application we therefore study the impact of different showers on the top- and bottom-mass effects with respect to the heavy-top approximation at the $13$\,TeV \lhc{}. For this purpose, \fig{fig:me} shows the ratio of the \nlo{}\plus{}\ps{} computation including mass effects and the corresponding cross section in the heavy-top limit as a function of the transverse momentum of the Higgs for different Monte Carlos: \pythia{} (black, solid), \hwpp{} (red, dotted), \pysix{} $p_T$-ordered (blue, dashed with points), \pysix{} $Q$-ordered (green, dash-dotted with open boxes) and \hwsix{} (yellow, solid with filled boxes). We apply the \mstw{} \nlo{} \pdf{} set \cite{Martin:2009iq} with the corresponding value of the strong coupling constant. The shower scale has been chosen as $\Qsh=\mh/2$ in all cases, while for $\muF$ and $\muR$ we use the defaults specified in \sct{sec:run}. Clearly, the mass effects are hardly dependent on the specific Monte Carlo which is particularly evident at small ($\pt{}\lesssim 50$\,GeV) and large ($\pt{}\gtrsim 150$\,GeV) transverse momenta. Nevertheless, there are some visible differences in the intermediate region which consistently discriminate the {\tt Herwig} from the {\tt Pythia} showers. Overall, they are at most $5$\% though and therefore still moderate.

\begin{figure}[t]
	\begin{subfigure}[b]{0.5\textwidth}
         	\includegraphics[width=\textwidth]{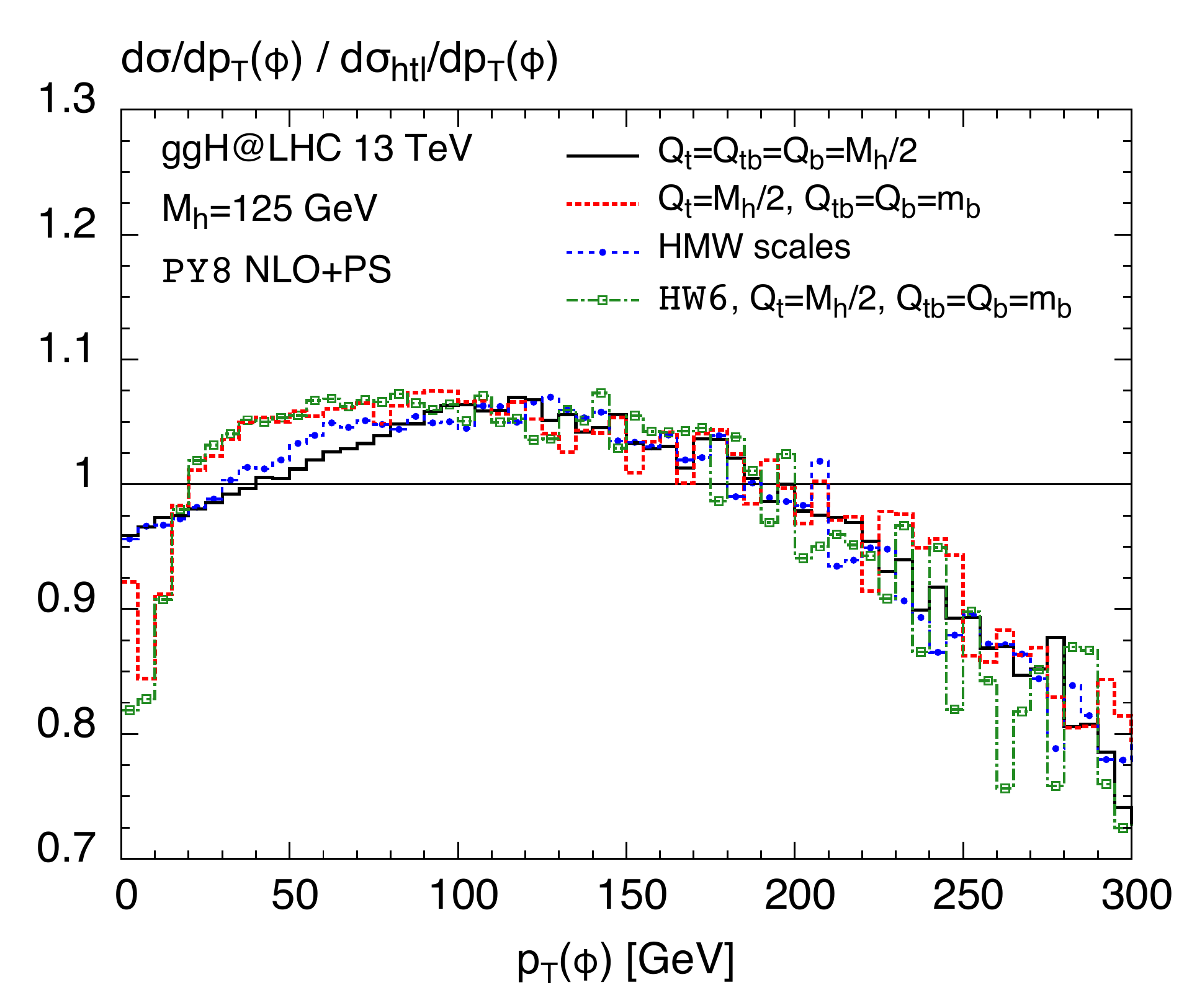}
	        \caption{}
            \label{fig:veto_NLO_mH_abs}
        \end{subfigure}%
	\begin{subfigure}[b]{0.5\textwidth}
         	\includegraphics[width=\textwidth]{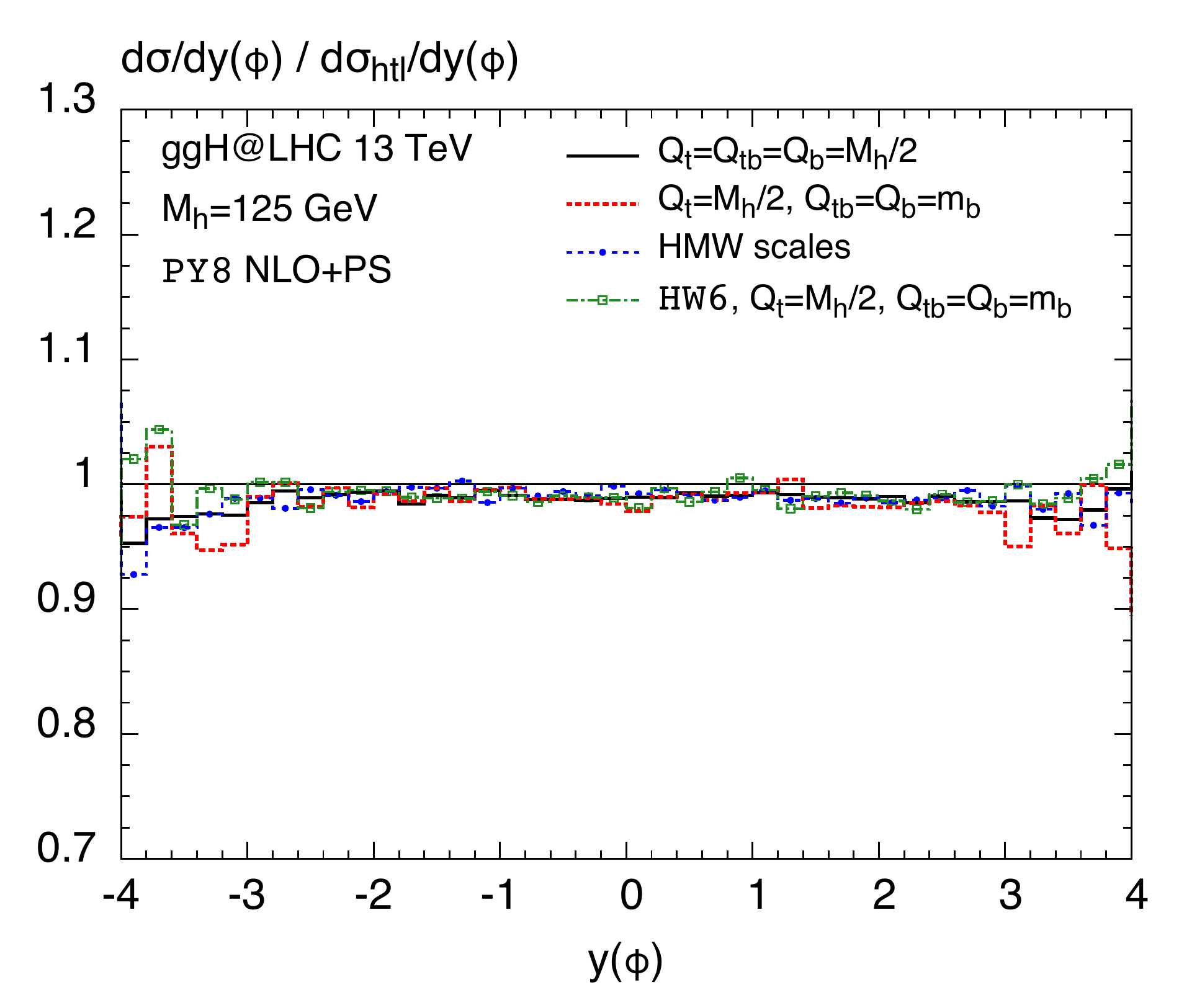}
		\caption{}
        \label{fig:veto_NLO_mH_rel}
        \end{subfigure}%
\begin{center}
        \parbox{.9\textwidth}{\caption{\label{fig:me2} (a) Same as \fig{fig:me}, but for different choices of the shower scales, see text for details; (b) corresponding plot for the rapidity distribution of the Higgs.
        }} \end{center}
\end{figure}

\fig{fig:me2} shows the effects of quark masses with respect to the heavy-top approximation as well, but for different choices of the associated shower scales. In all cases, the denominator and therefore the distribution in the heavy-top limit is computed with the respective scale of the top contribution $\Qsh=Q_t$. As we observed before the Monte Carlo dependence is quite small; therefore, we only consider the \pythia{} shower. For reference the black solid curve is the same as in \fig{fig:me} with $\Qsh=\mh/2$ for all contributions. We compare it to the scales choices proposed in \citere{Grazzini:2013mca}, which imply setting the shower scale of the bottom contribution (including the interference) to the bottom mass following \eqn{eq:twoscales} (red dotted curve). For the blue dashed curve with points we chose the {\abbrev HMW} scales determined in \citere{Harlander:2014uea} which can be found in Table 1 of that paper, applying a three-scale approach according to \eqn{eq:threescales} ($Q_t=49$\,GeV, $Q_{tb}=34$\,GeV and $Q_b=23$\,GeV). The green dash-dotted curve with open boxes serves mostly for comparison with previous \hwsix{} results \cite{Frixionetalk} which were computed with the scales of \citere{Grazzini:2013mca} as well. 

For the \pt{} distribution in \fig{fig:me2}\,(a), the change of the scale of the bottom contribution to $Q_b=m_b$ has a significant impact on the mass effects at small and intermediate transverse momenta. It develops an extremely steep drop at small transverse momenta which due to unitarity affects also the intermediate \pt{}-range in the opposite direction. The benefit of the usage of such a low scale is clearly disputable. While the \hwsix{} curve agrees rather well with previous result of \citere{Frixionetalk} becoming flat for $\pt\lesssim 5$\,GeV, the \pythia{} curve develops a steep increase in this region. This signals a significant Monte Carlo dependence at very small \pt{} which is not observed for larger $Q_b$ scales. Furthermore, the rigorously low value also poses a technical problem in the code regarding the fact that the default shower scale choice in \madmc{}, as explained in \sct{sec:scales}, is a distribution. In order to solve this problem, we had to use a fixed value of $\Qsh{}=Q_b$ by setting ${\tt frac\_low}={\tt frac\_upp}=Q_b/\mh{}$ and ${\tt scaleMClow}={\tt scaleMCdelta}=0$.

\begin{figure}[t]
	\begin{subfigure}[b]{0.5\textwidth}
         	\includegraphics[width=\textwidth]{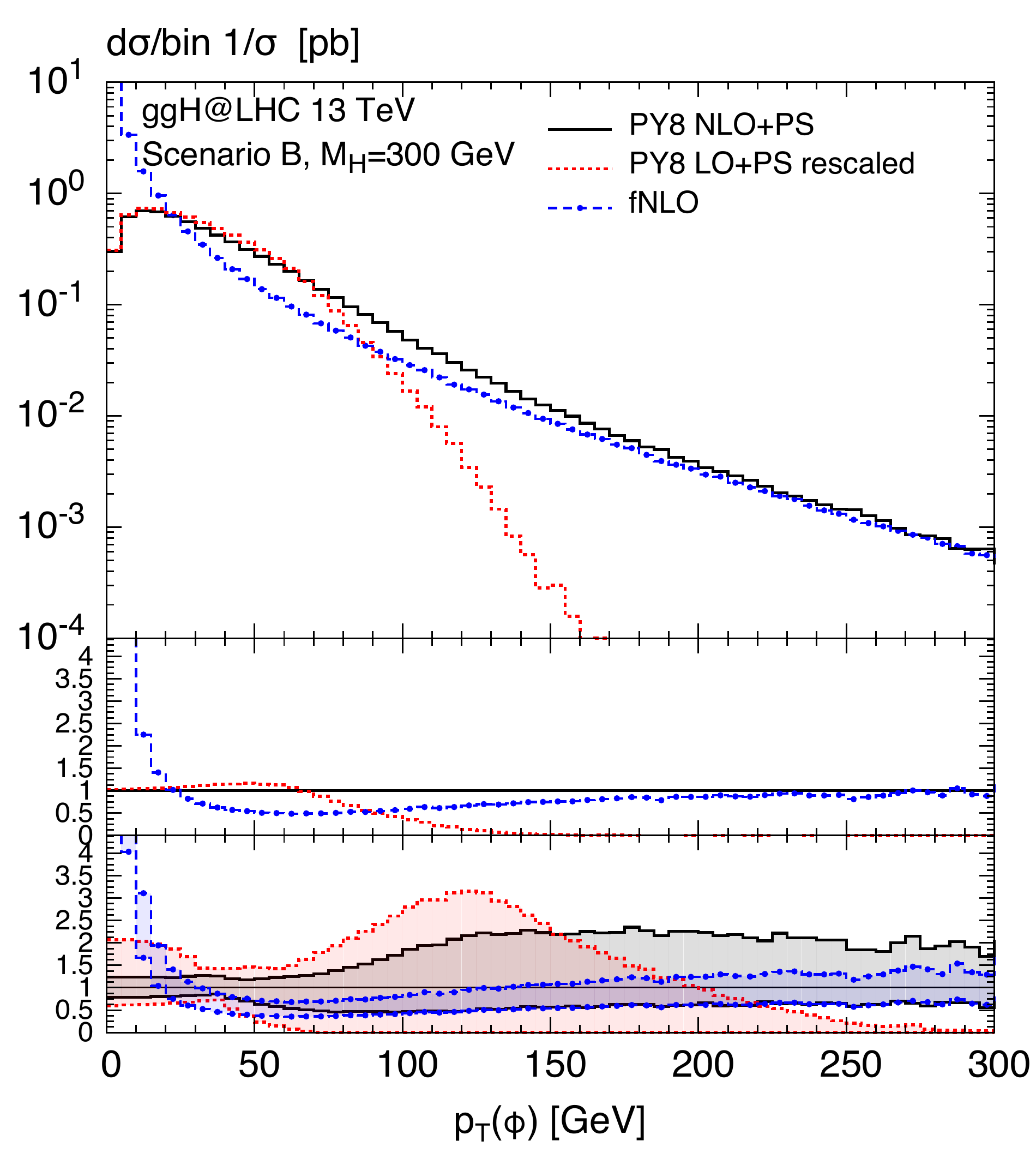}
	        \caption{}
            \label{fig:veto_NLO_mH_abs}
        \end{subfigure}%
	\begin{subfigure}[b]{0.5\textwidth}
         	\includegraphics[width=\textwidth]{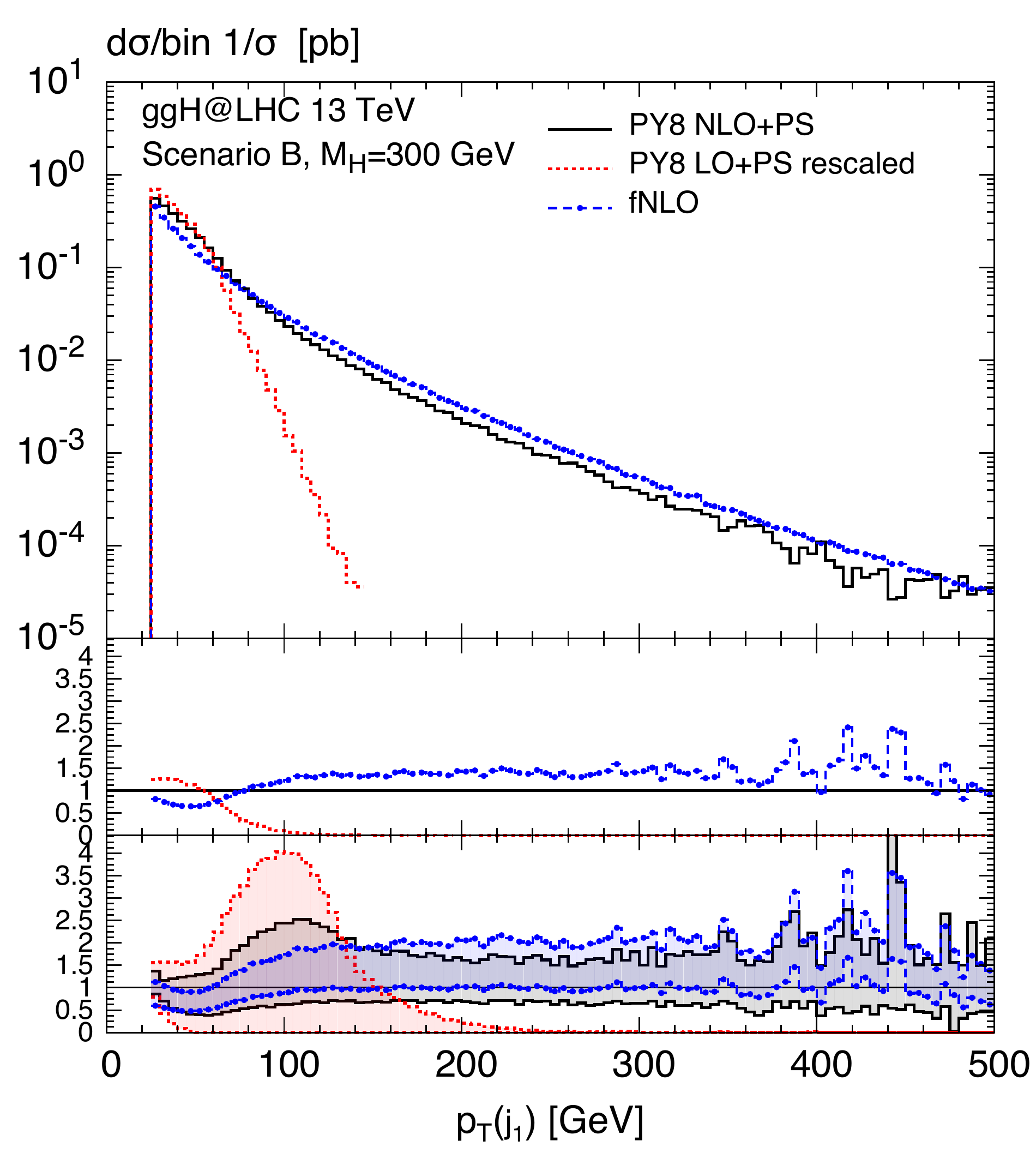}
		\caption{}
        \label{fig:veto_NLO_mH_rel}
        \end{subfigure}%
\begin{center}
        \parbox{.9\textwidth}{\caption{\label{fig:2HDM} Transverse momentum distribution of (a) the heavy Higgs boson and (b) the associated hardest jet computed in a bottom dominated scenario of the \thdm{} (see text for details). Graphical notation is the following: black solid curve shows \pythia{} at \nlo{}\plus{}\ps{}, red dotted curve is the same at \lo{}\plus{}\ps{} (normalized the \nlo{}) and the blue dashed one with points corresponds to the fixed order curve at \nlo{}.
        }} \end{center}
\end{figure}

Considering the {\abbrev HMW} scales, the scale of the bottom contribution is not chosen at such low values. We find that the mass effects in this case (blue dashed line with points) are rather similar to the ones where all scales are set to $\mh{}/2$ (black solid line), although the individual {\abbrev HMW} scales being quite different from this value. Looking at the rapidity distribution in \fig{fig:me2}\,(b), on the other hand, we observe the expected feature of being essentially insensitive to any choice of the respective shower scales. We shall note at this point that simply due to their inclusion in the default analysis we were able to produce a large number of further observables at no additional computing cost.

\begin{figure}[t]
	\begin{subfigure}[b]{0.5\textwidth}
         	\includegraphics[width=\textwidth]{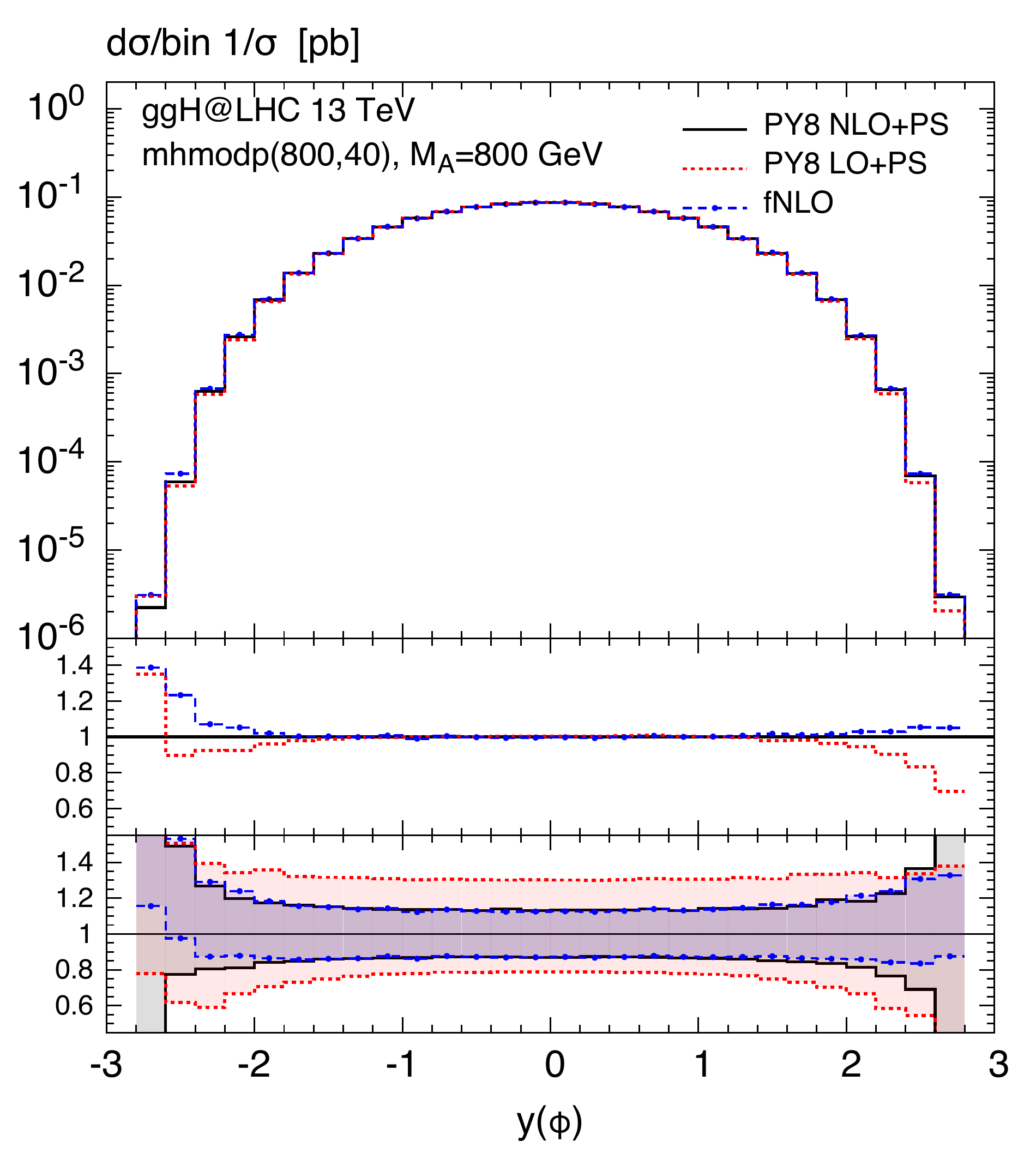}
	        \caption{}
            \label{fig:veto_NLO_mH_abs}
        \end{subfigure}%
	\begin{subfigure}[b]{0.5\textwidth}
         	\includegraphics[width=\textwidth]{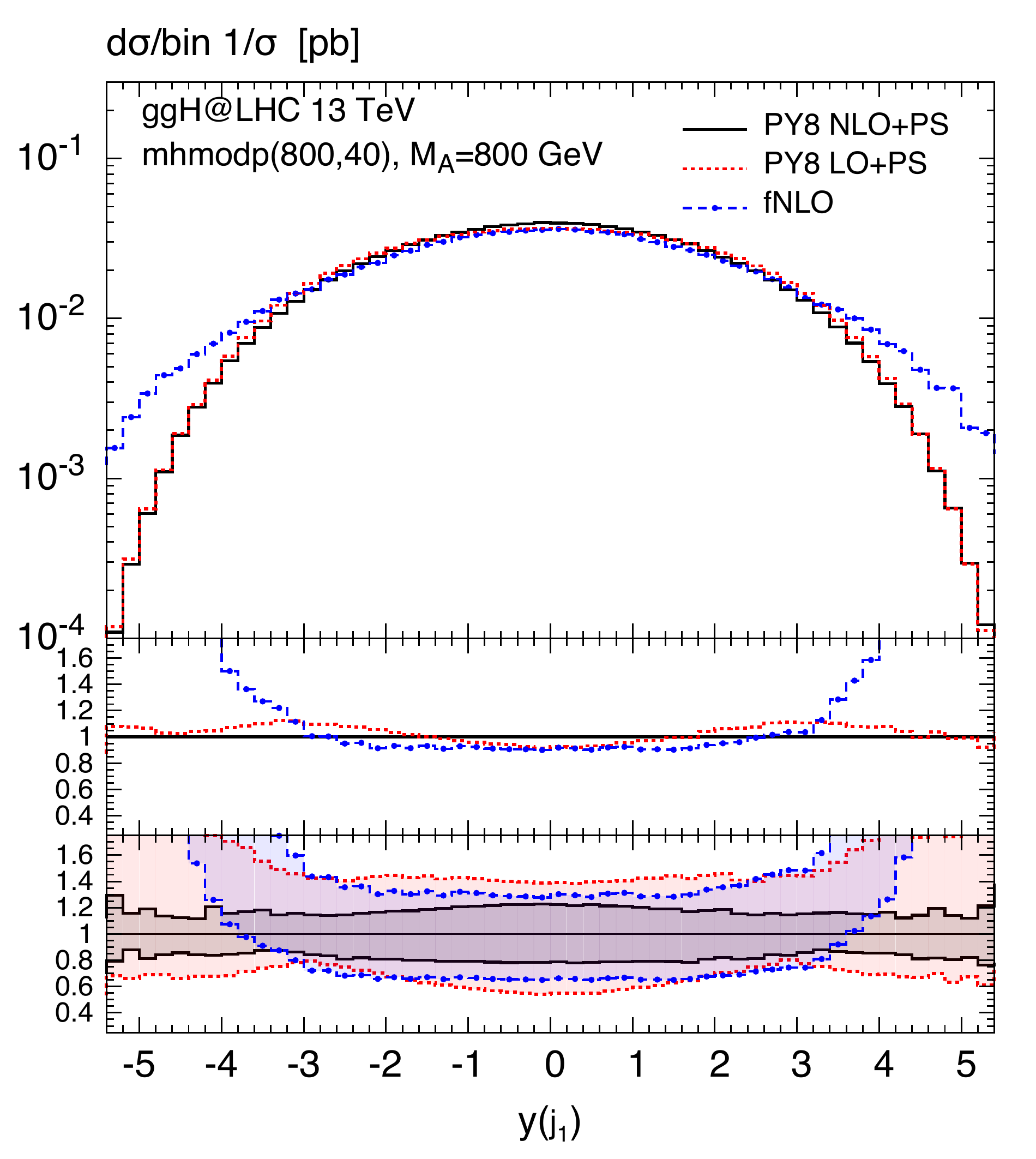}
		\caption{}
        \label{fig:veto_NLO_mH_rel}
        \end{subfigure}%
\begin{center}
        \parbox{.9\textwidth}{\caption{\label{fig:MSSM} Rapidity distribution of (a) the pseudo-scalar Higgs boson and (b) the associated hardest jet computed in the \mhmodp{} scenario \cite{Carena:2013qia} of the \mssm{} with $M_A=800$\,GeV und $\tan\beta=40$. Graphical notation is the same as in \fig{fig:2HDM}. All curves are normalized so that their bins add up to one.
        }} \end{center}
\end{figure}

To demonstrate the range of applicability of \sushimc{}, we consider two realistic \bsm{} scenarios in \fig{fig:2HDM} and \fig{fig:MSSM}: the heavy Higgs boson in Scenario B of \citere{Harlander:2013qxa} (a bottom dominated \thdm{} scenario); and the pseudo-scalar Higgs boson in the \mhmodp(800,40) \mssm{} scenario \cite{Carena:2013qia} defined in Table 2 of \citere{Harlander:2014uea}. The corresponding input files can be found in the folder $<$ggH-folder$>$/Cards. 

In \fig{fig:2HDM} we study the transverse momentum distributions of the Higgs and the hardest jet, while \fig{fig:MSSM} depicts their rapidity distributions. In both cases we apply the {\abbrev HMW} scales of \citere{Harlander:2014uea}. At low transverse momenta the \pt{} distributions have similar shapes comparing the red dotted (\lo{}\plus{}\ps{}) to the black solid curves (\nlo{}\plus{}\ps{}). This can be easily inferred from the first inset where all curves are normalized to the black solid line in the main frame. However, it is well known that at the \lo{}\plus{}\ps{} \pt{} distributions yield unphysical results for transverse momenta beyond the shower scales indicated by a steep drop. Note that both curves are normalized to the same (the \nlo{}) cross section. Continuing the comparison at hand, we observe a significant reduction of the scale uncertainties shown in the lower inset, where the bands are obtained by dividing the upper and lower bound of the respective cross section by the same central cross section as in the first inset. The uncertainties correspond to the independent variation of all unphysical scales ($\muR$, $\muF$, $Q_t$, $Q_b$, $Q_{tb}$) by a factor of two. Comparing \nlo{}\plus{}\ps{} to the \nlo{} fixed-order result denoted by f\nlo{}, we observe the expected matching towards large transverse momenta.

In order to compare shapes, the rapidity distributions in \fig{fig:MSSM} are normalized in a way that the sum of their bins yields one. We see that for the Higgs rapidity in \fig{fig:MSSM}\,(a) all curves agree extremely well in terms of shape up to the forward region in which, nevertheless, the deviations are still well within the respective uncertainty bands.\footnote{Note that the Higgs reaches its kinematical limit slightly before $|y(\phi)|\sim 3$ already.} For the rapidity distribution of the hardest jet the same is true for the two showered results, while the f\nlo{} distribution, on the other hand, agrees only in the central region $|y(j_1)|\lesssim 3$, but features a significant enhancement when the hardest jet is more forward. In this region the cross section will receive large effects of collinear radiation which renders the shower to yield the more reliable description.

\section{Conclusions}
\label{sec:conclusions}
In this article we presented the new tool \sushimc{} which is a link between \madmc{} and \sushi{} for the computation of Higgs cross sections in gluon fusion at hadron colliders. The code gives \nlo{}\plus{}\ps{} accurate results in the \sm{}, \thdm{} and \mssm{}. The inputs in the \mssm{} are conveniently controlled through a link to \fh{}. We discussed the specific treatment of the shower scale in \madmc{} and pointed out its special role in the context of gluon-induced Higgs production. In the phenomenological part we study the impact of different shower scale choices on the mass effects in the \sm{}. Furthermore, we studied results for \thdm{} and \mssm{} Higgs production as an application of \sushimc{}. The \sushimc{} script can be downloaded from \url{https://cp3.irmp.ucl.ac.be/projects/madgraph/wiki/HSushi\#no1}.

As an outlook, one may improve the \sm{} prediction for the Higgs production mode through gluon fusion by merging the \nlo{}\plus{}\ps{} cross section for $gg\rightarrow h$ in the full theory with higher multiplicities computed in the heavy-top effective field theory. This is certainly beyond the scope of the present paper and is left for a future publication.

\paragraph{Acknowledgements.}
We are indebted to Emanuele Bagnaschi, Stefano Frixione, Robert Harlander, Paolo Torrielli and Alessandro Vicini for several fruitful discussions and comments on the manuscript. We would like to thank Marco Zaro and Rikkert Frederix for their help on technical issues concerning \madmc{} and Stefan Liebler for providing the diagrams in \fig{fig:diag}. HM was supported by a Marie Curie Early Initial Training Network Fellowship of the European Community's Seventh Framework Programme under contract number (PITN-GA-2012-315877-MCnetITN). This research was supported in part by the Swiss National Science Foundation (SNF) under contract 200020-141360.

\appendix
\gdef\thesection{Appendix \Alph{section}}

\bibliographystyle{UTPstyle}
\bibliography{ggh_bib}

\end{document}